\documentclass{ws-ijmpe}

\begin{document}


\catchline{}{}{}{}{}

\title{Dense Matter and Neutron Stars in 
Parity Doublet Models
}

\author{\footnotesize S. Schramm}

\address{Frankfurt Institute for Advanced Studies, Johann Wolfgang Goethe Universit\"at\\Ruth-Moufang-Str. 1,
D-60438 Frankfurt, Germany
\\
schramm@fias.uni-frankfurt.de}

\author{V. Dexheimer}

\address{Physics Department, Gettysburg College, Gettysburg, Pennsylvania, 17325 USA\\
vantoche@gettysburg.edu}

\author{R. Negreiros}

\address{Frankfurt Institute for Advanced Studies, Johann Wolfgang Goethe Universit\"at\\Ruth-Moufang-Str. 1,
D-60438 Frankfurt, Germany
\\
negreiros@fias.uni-frankfurt.de}

\author{J. Steinheimer}

\address{Frankfurt Institute for Advanced Studies, Johann Wolfgang Goethe Universit\"at\\Ruth-Moufang-Str. 1,
D-60438 Frankfurt, Germany
\\
steinheimer@fias.uni-frankfurt.de}

\maketitle

\begin{history}
\received{(received date)}
\revised{(revised date)}
\end{history}

\begin{abstract}
We investigate the properties of dense matter and neutron stars.
In particular we discuss model calculations based on the parity doublet picture of
hadronic chiral symmetry. In this ansatz the onset of chiral symmetry restoration is
reflected by the degeneracy of baryons and their parity partners.
In this approach we also incorporate quarks as degrees of freedom to be able to study
hybrid stars. 
\end{abstract}

\section{Introduction}

One of the main and still unresolved problems in nuclear physics is the understanding of the properties and phase structure
of hot and dense matter. Here heavy-ion physics is an important experimental approach to extract information on very
hot and, depending on the beam energy, also dense matter. As a complementary approach the investigation of neutron stars 
can yield constraints on the properties of extremely dense and rather cold matter, where the temperature T is usually less than 1 MeV except
for the short period of the proto-neutron star, where T might reach 30 to 50 MeV.
Therefore both fields together can help to pin down, or at least constrain the equation of state and other properties of strongly interacting matter far beyond the
nuclear matter groundstate. One important aspect of QCD is its inherent chiral symmetry that characterizes the transition from low-density/temperature matter to matter at high values of $T$ and $\rho$ crossing from the symmetry broken to the restored phase.
There are different ways to construct a chirally symmetric hadronic model. One approach, given by the linear  $\sigma$ model and its many variations and extensions.  Here the masses of the baryons are largely generated by the coupling of baryons to the scalar field. Therefore, in the chirally restored phase the baryons become very light. Alternatively, in the so-called parity-doublet approach the baryonic degrees of freedom are augmented by the opposite-parity counterparts. \cite{Montvay1987} One possible state of such a particle is the N(1535) as parity partmer of the nucleon. In this formulation it is possible to define a mass term for the baryons without violating chiral symmetry. The chirally restored phase is characterized by the degeneracy of both parity partners. It has been shown that it is possible to describe saturating nuclear matter in such an approach.\cite{DeTar1989,Zschiesche:2006zj,Dexheimer2008,Dexheimer2008a} In the following we discuss results based on the latter approach, studying hadronic SU(2) and SU(3) models as well as an extended description including quark degrees of freedom.
We present results for the phase structure of the models as well as the properties of compact stars.

\section{The Hadronic SU(2) Parity Model}

In the parity doublet model positive and negative parity states of the baryons
are arranged in doublets \cite{Montvay1987,DeTar1989}. 
The two components of the fields defining the respective parity partners, $\varphi_+$ and $\varphi_-$
transform in opposite way regarding chiral transformations:
\begin{eqnarray}
&\varphi'_{+R}  =  R \varphi_{+R} &~,~ \varphi'_{+L}
= L \varphi_{+L} \ \nonumber \\ 
&\varphi'_{-R} = L \varphi_{-R} &~, \varphi_{-L}
= R \varphi_{-L} ~.
\end{eqnarray}
Here, $L$ and $R$ denote the rotations in the left- and right handed subspaces.
This allows for a chirally invariant mass term in the
Lagrangian of the general form:
\begin{eqnarray}
&m_{0}( \bar{\varphi}_- \gamma_{5} \varphi_+ - \bar{\varphi}_+
      \gamma_{5} \varphi_- ) =  \nonumber \\
&m_0 (\bar{\varphi}_{-L} \varphi_{+R} -
        \bar{\varphi}_{-R} \varphi_{+L} - \bar{\varphi}_{+L} \varphi_{-R} +
        \bar{\varphi}_{+R} \varphi_{-L}) ,
\end{eqnarray}
where $m_0$ represents a mass parameter. The coupling terms of the scalar field to the nucleonic states read
\begin{equation}
{\cal L_{BS}}  =  
a \bar{\varphi}_+ \left(\sigma + i \gamma_5 \vec{\tau}
  \cdot\vec{\pi}\right) \varphi_+
+ b \bar{\varphi}_-\left(\sigma - i \gamma_5 \vec{\tau}
  \cdot\vec{\pi}\right) \varphi_- .
\end{equation}
As both states in the doublet mix one  has to diagonalize the states as shown in \cite{Montvay1987}.
Fixing the coupling constants to reproduce the vacuum masses of the nucleons the resulting effective masses of the diagonalized states read \cite{DeTar1989} :
\begin{eqnarray}
{M_N^*}_\pm&=&\sqrt{\left[\frac{(M_{N_+}+M_{N_-})^2}{4}-m_0^2\right]\frac{\sigma^2}{\sigma_0^2}+m_0^2}\nonumber\\
&\pm&\frac{M_{N_+}-M_{N_-}}{2}\frac{\sigma}{\sigma_0}\ ,
\end{eqnarray}
where $\sigma_0$ denotes the vacuum value of the scalar field. As one can see from this expression both states are degenerate in the 
limit of a vanishing scalar field.

Doing an exhaustive parameter scan and constraining the saturation density $\rho_0$ to be 
about $0.15 / fm^3$ and the binding energy per nucleon $B/A$ to be around 16 MeV 
the resulting compressibilities $\kappa$ of nuclear matter are shown in Figure 1 as function of the mass parameter $m_0$.\cite{Dexheimer2008}
One can observe a strong dependence on this parameter. In order to have a reasonable value of $\kappa$ below 300 MeV a rather high value of $m_0$ around 840 MeV has to be chosen.
\begin{figure}[th]
\centerline{\includegraphics[width=6.5cm]{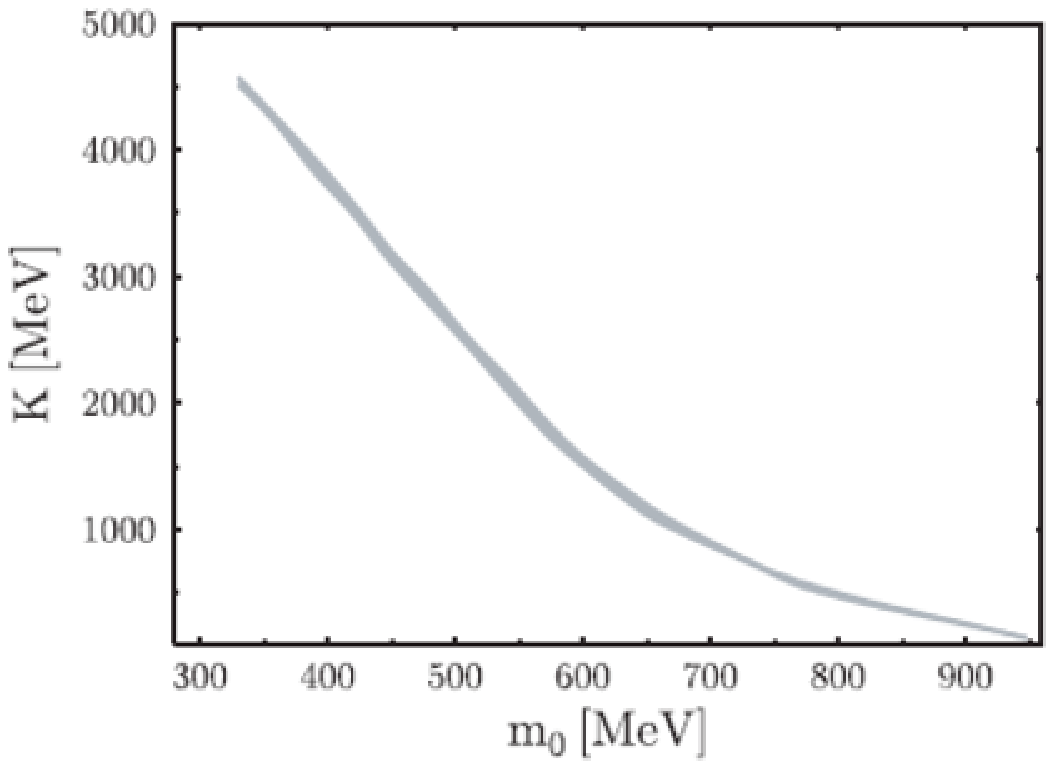}\includegraphics[width=7cm]{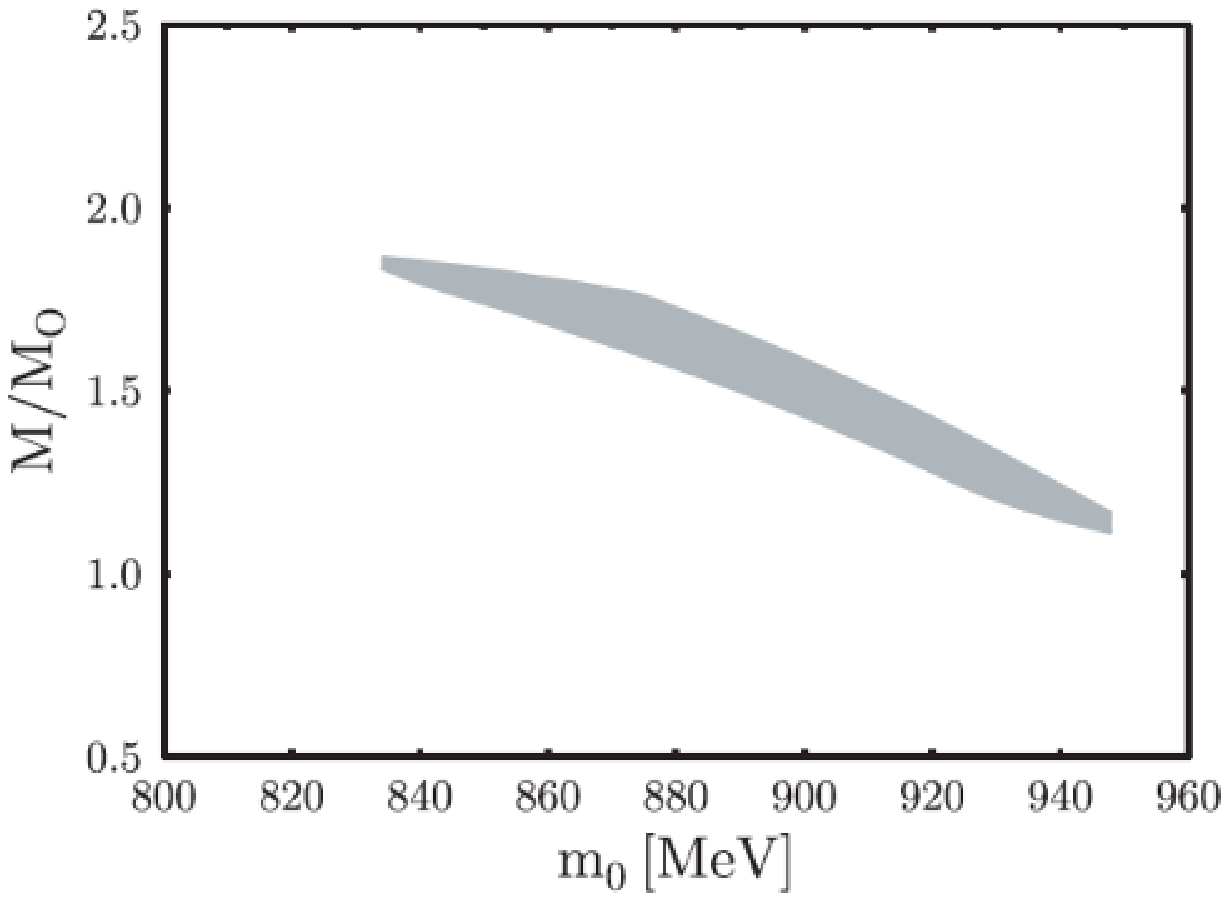}}
\vspace*{4pt}
\caption{Left Panel: Compressibility $\kappa$ as function of mass parameter $m_0$ given the mass of the nucleonic parity partmer $M_{N-} = 1535 $MeV.
Right Panel: Corresponding maximum neutron star masses in units of the solar mass.}
\end{figure}

Using the precious parameter scan with the same constraints on $\rho_0$ and $B/A$ and additionally demanding
that $\kappa$ has a value below 400 MeV the Tolman-Oppenheimer-Volkoff equation is solved for all acceptable parameter sets,
subsequently determining the maximum star  masses. The result of this calculation can be seen in Figure 1 (right panel). 

As the identification of the nucleonic parity partner is unclear one can study the model for 
different values of $M_{N-}$. Below a value of about 1380 MeV the parity partner states get populated in the core of the neutron star.
\begin{figure}[th]
\centerline{\includegraphics[width=6cm]{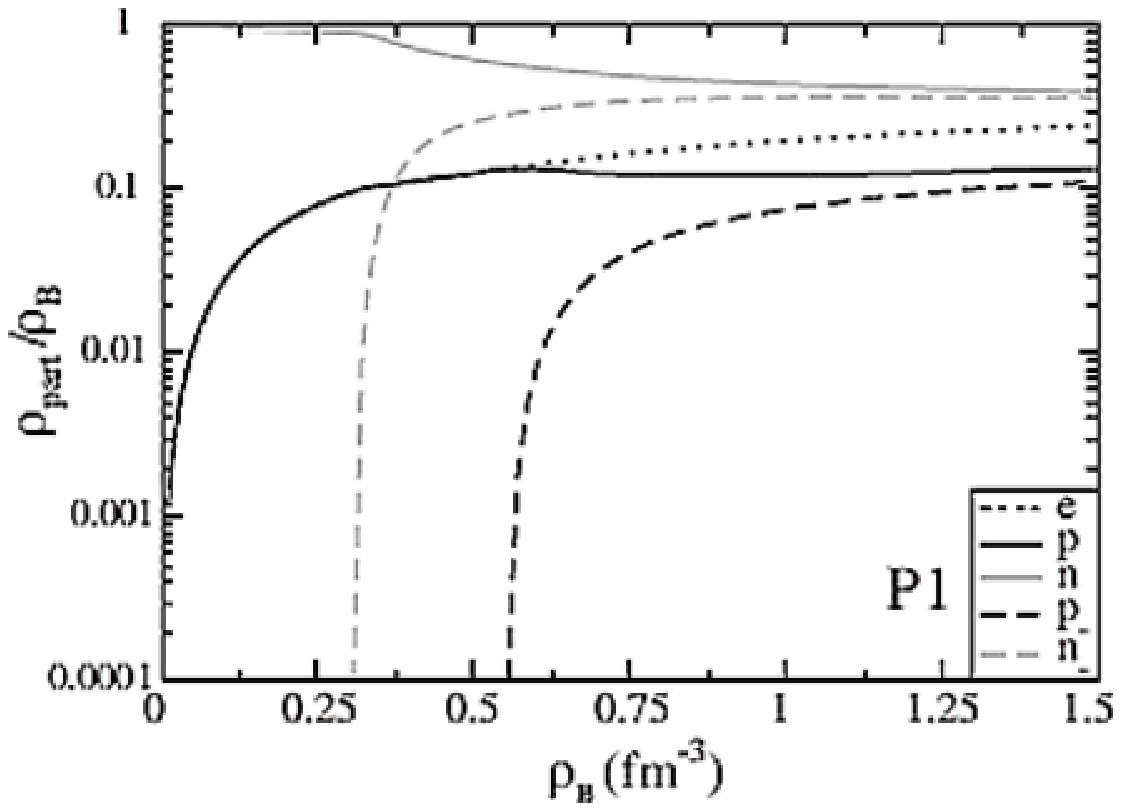}
\includegraphics[width=6cm]{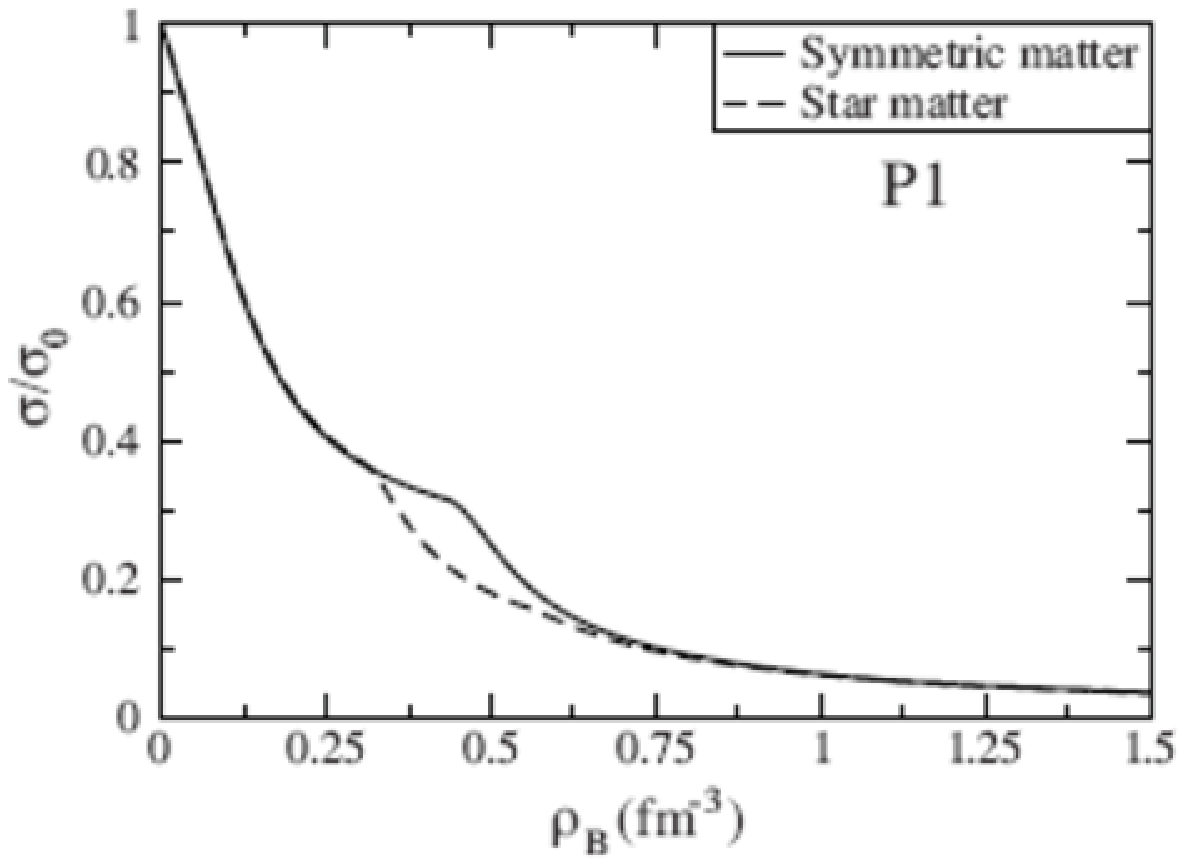}}
\vspace*{4pt}
\caption{Left Panel:Normalized particle densities as function of baryon density for star matter with a mass of the nucleonic parity partner of  $M_{N-} = 1200 $MeV (parameter set P1 in Ref. \protect\cite{Dexheimer2008a}).
Right Panel: Scalar field $\sigma$ as function of density}
\end{figure}

Setting the mass value to 1200 MeV (Fig. 2) a substantial fraction of the baryons inside the star are given by the parity partners that start to appear at
about $ 2\rho_0$ in the case of n$_-$ and $ 3\rho_0$ for p$_-$.\cite{Dexheimer2008a} One can observe that for large baryon densities the respective densities of the
nucleons and the parity partners become equal, which is a consequence of the degeneracy of the two states in the chirally restored phase.

This restoration of chiral symmetry can be seen in the behavior of the scalar field as it is shown in the right panel of Fig. 2. 
Here also the difference of the drop of the scalar field value
in the case of isospin symmetric and star matter is shown. As it is generally the case the transition becomes weaker and shifts to lower densities for star matter.
This is due to the breaking of the isospin symmetry in the stellar environment. Thus, the parity partner of the neutron is populated earlier, but as this onset of population affects only the n$_-$ compared to the simultaneous appearance of the n$_-$ and p$_-$ in the case of isospin symmetric matter the effect on the scalar field is reduced.

\section{The SU(3) Parity Model}

Extending the model to flavor SU(3) in a nonlinear realization of chiral symmetry we follow the discussion in \cite{Nemoto1998}. In general the formalism is analogous to the SU(2) case
(see Ref. \cite{su3parity} for a detailed discussion). The baryonic Lagrangian reads
\begin{equation}
{\cal L_B} = \sum_i (\bar{B_i} i {\partial\!\!\!/} B_i)
+ \sum_i  \left(\bar{B_i} m^*_i B_i \right) +
\sum_i  \left(\bar{B_i} \gamma_\mu (g_{\omega i} \omega^\mu +
g_{\rho i} \rho^\mu + g_{\phi i} \phi^\mu) B_i \right)
\label{lagrangian2}
\end{equation}
with couplings to the vector fields $\omega, \rho$, and $\phi$.
The effective masses of the baryon octet and the corresponding parity partners are given by the expression
\begin{equation}
m^*_{i\pm} = \sqrt{ \left[ (g^{(1)}_{\sigma i} \sigma + g^{(1)}_{\zeta i}  \zeta )^2 + (m_0+n_s m_s)^2 \right]}
\pm g^{(2)}_{\sigma i} \sigma \pm g^{(1)}_{\zeta i} \zeta
\end{equation}
including a SU(3) symmetry breaking term proportional to the number of strange quarks $n_s$ of the baryon. $m_s$ has been fixed to a value of
150  MeV. The index  $i$ denotes the baryon and $\pm$ are the positive and negative parity partners as in the SU(2) case. Note that in the SU(3) formulation there
are two scalar fields $\sigma$ and $\zeta$, which are effectively the counterparts to the QCD scalar non-strange and strange quark condensates.
The scalar meson interaction driving the spontaneous breaking of chiral symmetry
can be written in terms of SU(3) invariants 
$I_1 = Tr(\Sigma)  ~,~ I_2 = Tr(\Sigma^2) ~,~ I_3 = det(\Sigma) ~,~ I_4 = Tr (\Sigma^4) $, where $\Sigma$ denotes the multiplet of scalar mesons:\cite{Papazoglou:1997uw}
\begin{equation}
V = V_0 + \frac{1}{2} k_0 I_2 - k_1 I_2^2 - k_2 I_4 + k_3 {\rm ln}(I_3)~,
\end{equation}
where $V_0$ is fixed by demanding a vanishing potential in the vacuum.
The explicit symmetry breaking term that generates the correct pion and kaon masses with their corresponding decay constants
can be written as \cite{Papazoglou:1997uw}
\begin{eqnarray}
&L_{SB}= m_\pi^2 f_\pi\sigma+\left(\sqrt{2}m_k^ 2f_k-\frac{1}{\sqrt{2}}m_\pi^ 2 f_\pi\right)\zeta.\nonumber&\\&
\end{eqnarray}

Fitting the coupling constants to a reasonable nuclear matter ground state as in the previous section we solve the equations of motion for the various fields.
In a further extension we include quark degrees of freedom and the Polyakov loop field $\Phi$ as order parameter for the deconfinement transition \cite{Ratti:2005jh,Fukushima:2003fw}. Here
we follow the formulation as outlined in Ref. \cite{Steinheimer:2010ib}. The quarks couple to the condensates and attain an effective mass:
\begin{equation}
m_{q}^*=g_{q\sigma}\sigma+ m_{0q},~~~
m_{s}^*=g_{s\zeta}\zeta+m_{0s},
\end{equation}
with couplings $g_{q\sigma}=g_{s\zeta}= 4.0$. $m_{0q}$ and $m_{0s}$ describe
the current quark masses as well as contributions from the slowly varying gluon condensate \cite{su3parity}.
The potential of the Polyakov loop effectively represents the contribution of the
gluons to the thermodynamical quantities of the system like pressure and energy density.
Here we adopt the ansatz proposed in \cite{Ratti:2005jh}:
\begin{equation}
	U = -\frac12 a(T)\Phi\Phi^*
	 + b(T)ln[1-6\Phi\Phi^*+4(\Phi^3\Phi^{*3})-3(\Phi\Phi^*)^2]
\end{equation}
 with $a(T)=a_0 T^4+a_1 T_0 T^3+a_2 T_0^2 T^2$, $b(T)=b_3 T_0^3 T$, where $T_0$ sets the scale for the first-order deconfinement
 phase transition in the quenched limit of only gluons.\\ Finally, we introduce an excluded volume correction for the hadrons 
 \cite{Rischke:1991ke,Cleymans:1992jz}, which leads to the suppression
 of hadronic degrees of freedom relative to quarks at higher densites as described in \cite{su3parity}.
 
Using the parameter set for a mass of the negative parity state of the nucleon $M_{N-} = 1535 $MeV and assuming, for simplicity, the same mass splitting of all baryon doublets as the nucleonic
one, as the  possible attribution of observed states is less clear for the hyperons, \cite{su3parity}  we study the phase transition behavior of the model as function of chemical potential and
temperature. The result is shown in Fig. 4. The two plots are calculated for different values of the temperature parameter in the Polyakov loop potential
$T_0 = 270$ and $220$ MeV, respectively. The lower band indicates the chiral transition region during which the value of $\sigma$ drops from 0.8 to 0.2 of its vacuum value. The upper band corresponds to the deconfinement transition connected to the value $\Phi$ of the Polyakov loop changing from 0.2 to 0.8. The two first-order transition lines for the liquid-gas and chiral transition are shown as solid lines. Both stop at a critical end-point with values (in units of MeV)
of 
$(T,\mu) = (17,903)$ and $(58,1200)$, respectively.
\begin{figure}[th]
\centerline{\psfig{file=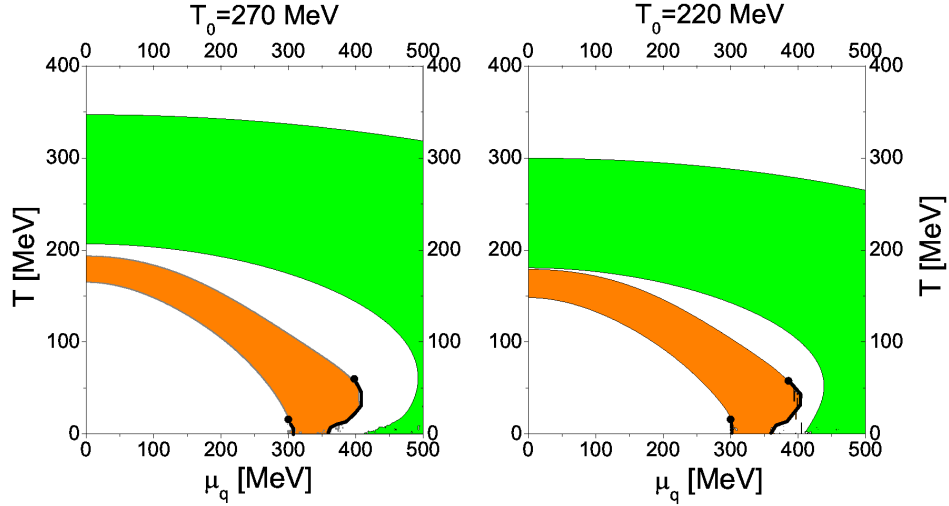,width=14cm}}
\vspace*{8pt}
\caption{Phase transition diagram as function of temperature $T$ and quark chemical potential $\mu_q = \mu_B/3$. The two panels are the results
for different parameter values for $T_0$ of 270 (left) and 220 MeV (right). The lower band shows the range where  the scalar field has a value between
20 and 80 percent of its vacuum value. Analogously,  the upper band indicates the range of temperatures and chemical potentials for which the Polyakov loop
field lies between 0.2 and 0.8. The solid lines show the first-order liquid-gas and chiral phase transitions. In case of the lower value of 
$T_0$ the baryochemical potential of the critical end point moves from 1200 to 1150 MeV.}
\end{figure}

\begin{figure}[th]
\centerline{\includegraphics[width=7cm]{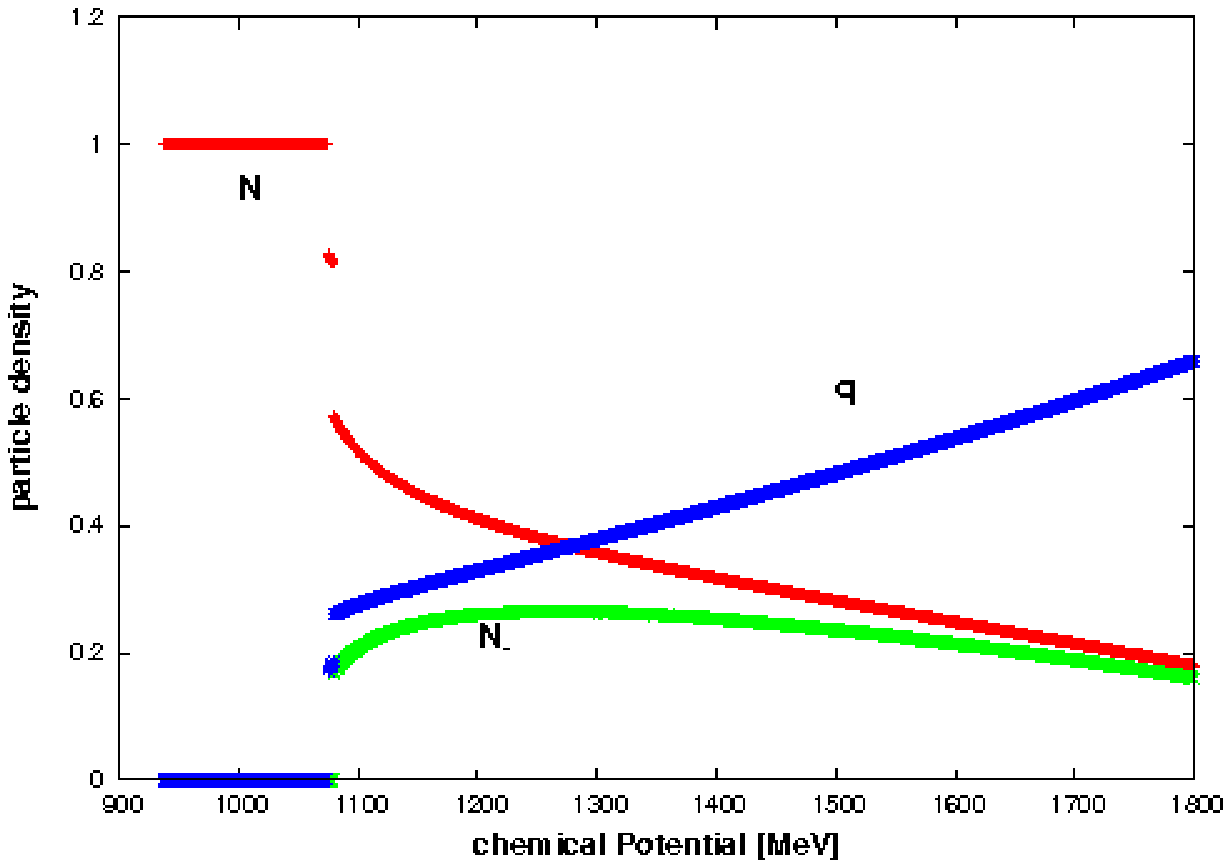}\includegraphics[width=7cm]{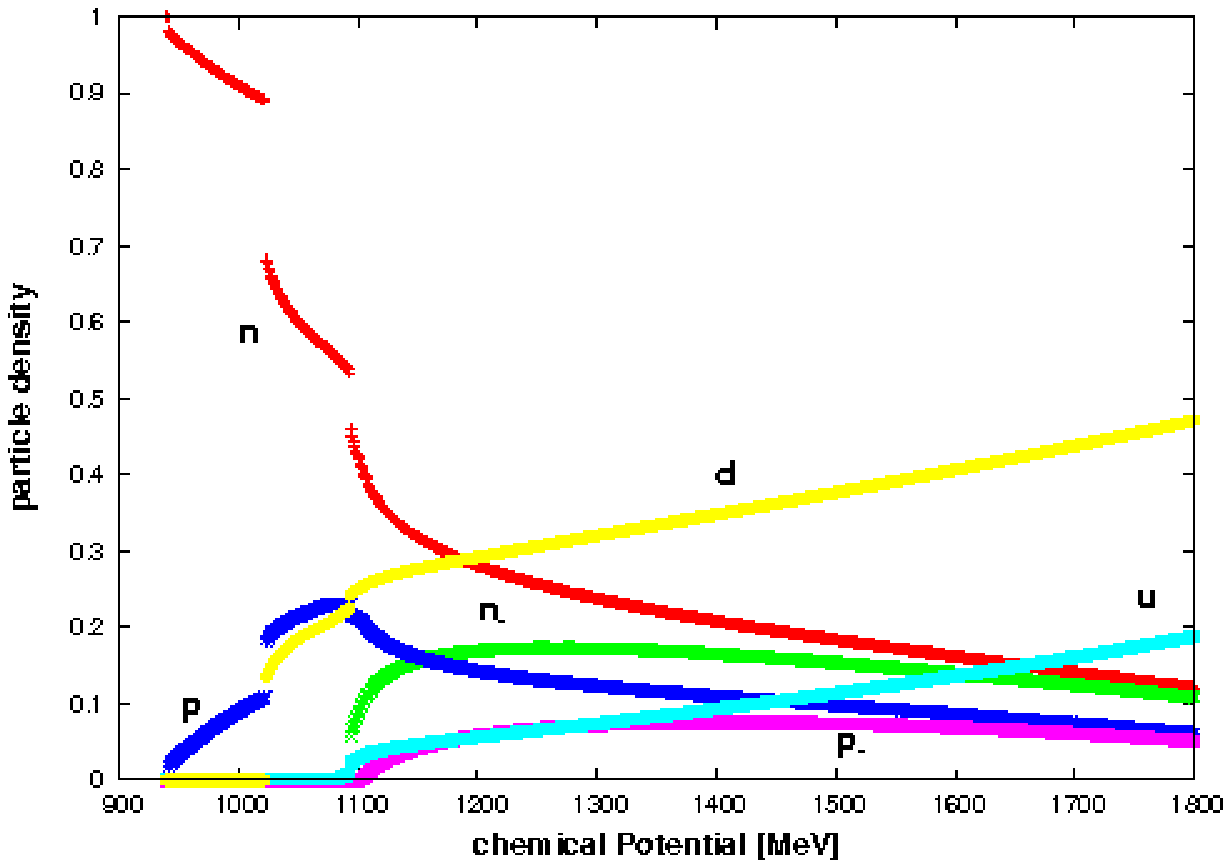}}
\vspace*{5pt}
\caption{Normalized baryon number densities for the different particles as function of chemical potential for isospin symmetric matter (left panel)
and beta-equlibrated stellar matter (right panel) at $T=0$.}
\label{popsu3}
\end{figure}

In order to study the particle species that are populated the densities of the various degrees of freedom as function of 
chemical potential are shown in Fig. \ref{popsu3}. The results in the left panel are obtained for zero temperature and isospin-symmetric matter.
One can see the onset of the occurrence of quarks and the nucleonic parity partners at slightly higher chemical potential,
such that there actually is a two-step first-order transition. In contrast to the hadronic model the parity partner appears earlier
driven by the appearance of the quarks and the connected drop of the scalar condensate.  
A similar, but more involved behavior can be seen in the right panel for star matter. First the down quark appears in a first-order transition,
then, in a second transition the positively charged up quark and the nucleonic parity partners are populated. The hyperonic states and their parity partners are suppressed by the occurrence of the
quarks and appear only at much higher chemical potential.

\begin{figure}[th]
\centerline{\includegraphics[width=8cm]{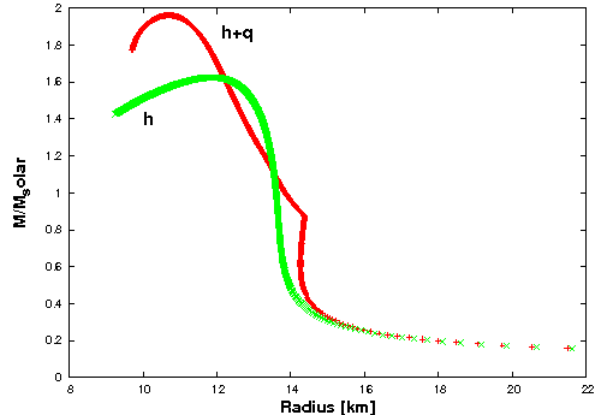}}
\vspace*{4pt}
\caption{Mass-radius diagram for compact stars in the SU(3) parity model including hadrons (h), and quarks and hadrons (h+q).}
\label{starsu3}
\end{figure}

Using the thus obtained equation of state we solve the TOV equations varying the central pressure of the star.
The resulting star sequences are given in Figure \ref{starsu3}. The result for the purely hadronic model with a maximum mass of
about 1.6 solar masses and a radius of 12 km. Including quarks shifts the maximum mass to 1.96 solar masses with a reduced radius of
10.7 km, which is in agreement with the rather accurate measurement of the $1.97\pm0.4$  solar mass pulsar PSR J1614-2230. \cite{197} The kink in this sequence indicates the onset of a quark core in the star, i.e. the transition from a neutron to a hybrid star. The fact that the masses for both cases below that value are not identical originates from the fact that only in the case of the full calculation with quarks the excluded volume corrections are taken into account. 

Summarizing, we discussed SU(2) and SU(3) parity doublet models for hadronic matter with and without quark degrees of freedom. We could reproduce nuclear matter saturation properties reasonably well. Depending on the mass of the parity partners the model has a first-order chiral phase transition at high densities for larger masses and a crossover for smaller masses, respectively. Including quark degrees of freedom triggers a first-order transition for temperatures below $\approx 60\,$MeV. The calculation of static hybrid stars yields maximum star masses of 2 solar masses in agreement with observation. 

\section*{Acknowledgements}

We acknowledge access to the computer facilities of the CSC Frankfurt. R. N. and J. S. acknowledge financial support from the LOEWE program HIC for FAIR.


\begin{thebibliography}{99}
\bibitem{Montvay1987}
I. Montvay, {\it Phys. Lett. } {\bf B. 199} (1987) 89. 
\bibitem{DeTar1989}
C. DeTar, T. Kunihiro, {\it Phys. Rev.} {\bf D 39} (1989) 2805.
\bibitem{Zschiesche:2006zj}
 D. Zschiesche, L.  Tolos, J.  Schaffner-Bielich,  R. D. Pisarski, {\it Phys. Rev.}  {\bf C 75}, 055202 (2007).
\bibitem{Dexheimer2008}
V. Dexheimer, S. Schramm, D.  Zschiesche, {\it  Phys. Rev.} {\bf C 77} (2008) 025803. 
\bibitem{Dexheimer2008a}
V. Dexheimer, G. Pagliara, L. Tolos, J. Schaffner-Bielich, S. Schramm, {\it Eur. Phys. J.} {\bf A38} (2008) 105. 
\bibitem{Nemoto1998}
Y. Nemoto, D. Jido, M. Oka  {\it Phys. Rev.} {\bf D 57} (1998) 4124.
\bibitem{su3parity} 
J. Steinheimer, S. Schramm, H. St\"ocker, preprint arXiv:1108.2596  [hep-ph].
\bibitem{Papazoglou:1997uw}
  P.~Papazoglou, S.~Schramm, J.~Schaffner-Bielich, H.~St\"ocker and W.~Greiner,
  {\it Phys.\ Rev.\  } {\bf C 57}, (1998) 2576 .
\bibitem{Ratti:2005jh}
  C.~Ratti, M.~A.~Thaler and W.~Weise,
  {\it Phys.\ Rev.\  } {\bf D 73}, (2006) 014019.
\bibitem{Fukushima:2003fw}
K.~Fukushima,
{\it  Phys.\ Lett.\  }  {\bf B591}, (2004) 277.
\bibitem{Steinheimer:2010ib}
  J.~Steinheimer, S.~Schramm, H.~St\"ocker,
{\it Phys. Lett. } {\bf B 696} (2011) 257.
\bibitem{Rischke:1991ke}
  D.~H.~Rischke, M.~I.~Gorenstein, H.~Stoecker and W.~Greiner,
  {\it Z.\ Phys.\  } {\bf C51}, (1991) 485.

\bibitem{Cleymans:1992jz}
  J.~Cleymans, M.~I.~Gorenstein, J.~Stalnacke and E.~Suhonen,
{\it  Phys.\ Scripta }{\bf 48}, (1993) 277.
\bibitem{197}
P. B. Demorest, T. Pennucci, S. M. Ransom, M .S. E. Roberts, J. W. T.. Hessels, {\it Nature} {\bf 467},  (2010) 1081. 
\end{thebibliography}
\end{document}